\documentclass[aps,prl,twocolumn,superscriptaddress]{revtex4}
\usepackage{graphicx}

\begin{document}

\title{Topologically Nontrivial Bismuth(111) Thin Films Grown on Bi$_2$Te$_3$}

\author{Meng-Yu Yao}
\author{Fengfeng Zhu}
\author{Lin Miao}
\author{C. Q. Han}
\author{Fang Yang}
\author{D. D. Guan}
\affiliation{Key Laboratory of Artificial Structures and Quantum Control (Ministry of Education), Department of Physics and Astronomy, Shanghai Jiao Tong University, Shanghai 200240, China}
\author{C. L. Gao}
\author{Canhua Liu}
\author{Dong Qian}
\email{dqian@sjtu.edu.cn}
\author{Jin-feng Jia}
\affiliation{Key Laboratory of Artificial Structures and Quantum Control (Ministry of Education), Department of Physics and Astronomy, Shanghai Jiao Tong University, Shanghai 200240, China}
\affiliation{Collaborative Innovation Center of Advanced Microstructures, Nanjing 210093, China}

\date{\today}

\begin{abstract}
Using high-resolution angle-resolved photoemission spectroscopy, the electronic structure near the Fermi level and the topological property of the Bi(111) films grown on the Bi$_2$Te$_3$(111) substrate were studied. Very different from the bulk Bi, we found another surface band near the $\bar{M}$ point besides the two well-known surface bands on the Bi(111) surface. With this new surface band, the bulk valence band and the bulk conduction band of Bi can be connected by the surface states. Our band mapping revealed odd number of Fermi crossings of the surface bands, which provided a direct experimental signature that Bi(111) thin films of a certain thickness on the Bi$_2$Te$_3$(111) substrate can be topologically nontrivial in three dimension.
\end{abstract}

 \maketitle
Topological insulators (TIs) that possess the nontrivial topological surface states have been systematically studied in the last several years\cite{Bernevig2006,Konig2007,Fu2007,Hsieh2008,Zhang2009,Xia2009,Chen2009,Hasan2010a,Qi,Wang2012}. Bi$_{1-x}$Sb$_{x}$ alloy is the first experimentally realized three dimensional (3D) TI\cite{Hsieh2008}. Without Sb, pure bulk Bi is a semi-metal and famous for its novel surface states that is related to the very large Rashba-type spin orbital coupling (SOC)\cite{Hofmann}. Though Bi's surface states are very robust experimentally\cite{Jinarxiv, TianSrp}, bulk Bi is topologically trivial in theory\cite{Fu2007}. When alloyed with Sb, the band inversion occurs at the L point and Bi$_{1-x}$Sb$_{x}$ becomes a TI\cite{Fu2007,Hsieh2008}. The natural cleaving surface of Bi is the (111) surface. In Fig. 1(a), we illustrate the low energy bands of bulk Bi(111) near the Fermi level including the surface states according to the LDA calculations\cite{exp4,JapPRL}. Fermi level crosses both the bulk valence band ("BVB" in the figure) and the bulk conduction band ("BCB" in the figure) forming a semi-metallic state. Due to the Rashba-type SOC, there are two spin splitting surface bands ("S1" and "S2" in the figure). According to the Kramers theorem, S1 and S2 bands must be degenerate at the time-reversal invariant points, $\bar{\Gamma}$ and $\bar{M}$ points in the surface hexagonal Brillouin zone (BZ), protected by the time reversal symmetry. LDA calculations predicted that there should be even number of Fermi crossings of the surface bands, since bulk Bi is topologically trivial\cite{Fu2007}. However, so far the experimental findings are very complicated and unclear. The electronic structure of the bulk Bi(111) surface has been extensively studied by angle-resolved photoemission spectroscopy (ARPES)\cite{Hofmann,exp4,exp1,exp2,exp3,Ast2001,Ast2003,Ohtsubo2013}. Figure 1(b) shows the sketch of the previously reported experimental bands on the Bi(111) surface\cite{Hofmann,exp4}. The surface bands S1 and S2 do merge together into the bulk valence band at the $\bar{\Gamma}$ point, but they are not degenerate at the $\bar{M}$ point at all, which is totally in contrast to the LDA prediction. At the $\bar{M}$ point, the S2 band merges into the bulk valence band, but the S1 band is above the bulk valence band. Unfortunately, no bulk conduction band near the $\bar{M}$ point have been detected by ARPES on the bulk Bi samples. If the S1 band merges into the bulk conduction band, then Bi will be a topologically nontrivial semimetal just like Sb\cite{DavidScience}. Because of the missing experimental signature of the bulk conduction band and of the degeneracy of S1 and S2 bands in the $\bar{M}$ point, recent theoretical work claimed that the bulk Bi is topologically nontrivial\cite{Ohtsubo2013}.

\begin{figure}[b]
\center \includegraphics[width=8cm]{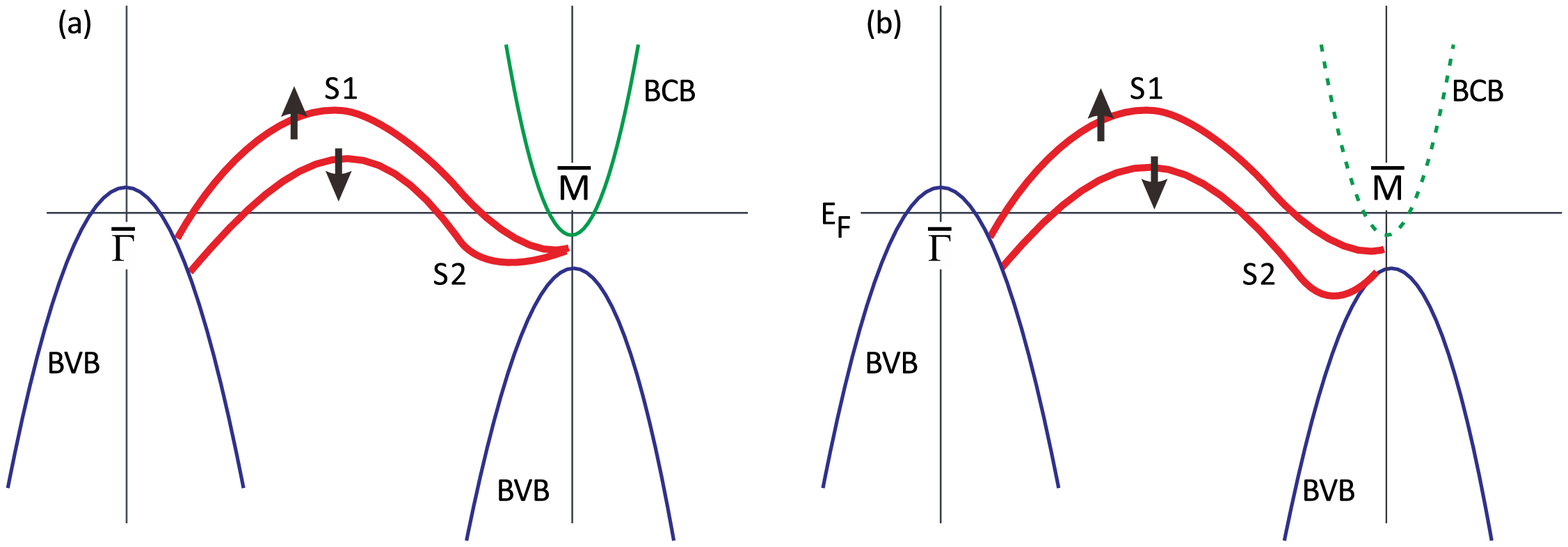} \caption{Sketch of the low energy bands of bulk Bi(111). (a) Bands according to the LDA calculations. (b) Bands according to the ARPES experiments. Red bands are the surface bands S1 and S2. Black arrows indicate the spin polarization. Electrons in two surface bands have opposite spin polarization. Blue bands are the bulk valence bands and green bands are the bulk conduction bands. Green dashed line means that it has not been detected in experiments.}
\end{figure}

On the other hand, high quality single crystalline Bi films with $<$111$>$ orientation can be grown by molecular beam epitaxy method\cite{Hirahara2006,JinPRL}. The band structure of Bi(111) films is very sensitive to the film thickness and the lattice constant\cite{JapPRL,KoroteevPRB,Miao_PRB,LiuPRL,HiraharaPRL,YangPRL}. For example, the ultra-thin Bi(111) films ($<$ 2nm) can host the quantum spin Hall state\cite{LiuPRL,HiraharaPRL,YangPRL}. Interestingly, recent transport measurements on the Bi(111) films and Bi nanoribbons revealed some experimental evidences of the existence of the topologically protected surface states\cite{JinPRL, Jinarxiv, TianSrp}. New theory was developed to show that the Bi(111) thin films could be a 3D TI like\cite{Jinarxiv}. ARPES is the direct method to check the topological property of a 3D TI\cite{Hasan2010a}. In the last decade, many ARPES experiments have been carried out on the Bi(111) films on Si(111)\cite{Hirahara2006,Hirahara2007,Takayama2011,Takayama2012,Takayama2014}. In the ultrathin regime, the hybridization between the surface states from the bottom and top surfaces and the bulk quantum states was observed\cite{Hirahara2006,Takayama2012}. In the thicker films, the low energy ARPES spectra as well as the Fermi surface topology become very similar to the results on the bulk Bi(111)\cite{Takayama2012}, which means that we can not make conclusion that whether the Bi(111) films are topologically trivial or not based on the reported ARPES band mapping on the Bi(111)/Si(111) films.

Very recently, high quality ultrathin Bi(111) films were also obtained on the Bi$_2$Te$_3$(111) substrate\cite{HiraharaPRL,YangPRL,Miao_PRB}. This new system provides us a new opportunity to explore the topological property of the Bi(111) films. In this work, we studied the low energy electronic structure of the Bi(111) films grown on the Bi$_2$Te$_3$(111) substrate using the high-resolution ARPES at low temperature (T=10 K). For the first time, we found three surface bands near the Fermi level. Our findings suggest that the Bi(111)/Bi$_2$Te$_3$(111) film of a certain thickness is topologically nontrivial.

ARPES experiments were carried out in Advanced Light Source Beamline 12.0.1 with the incident photons of from 28 to 46 eV. All the spectra were taken at 10 K using a Scienta analyzer with a base pressure of better than 2$\times$10$^{-11}$ Torr. The polycrystalline Au electronically contacted with samples was used as the reference of the Fermi level. The energy resolution is about 10 meV and the angular resolution is better than 1\% of the surface BZ. High quality Bi$_2$Te$_3$ bulk single crystals as well as thin films were used as the substrates. Bulk single crystals are grown by modified Bridgman method. Single crystals were cleaved \textit{in situ}  at 10 K, resulting in shiny and well-ordered (111) surfaces. 40nm Bi$_2$Te$_3$(111) thin films were grown by MBE on Si(111) wafer. Bi(111) films grow as the bilayer (BL) growth mode. The thickness of each BL is about 0.4 nm. In order to get very high quality Bi(111) films, we used a "two-step" growth method. First, the substrate was kept at 250K during the growth of the first 15 BLs according to our previous study\cite{Miao_PRB, Miao_PNAS}. ii) Secondly, we raised the substrate temperature to 420K to grow more BLs. The deposition rate of Bi was about 0.3 BL/min. Fig. 2(a) shows the RHEED pattern of the 30nm Bi(111) films. It is sharp and line-like, which means the high-crystalline-quality and very flat surface. High quality of the film's surface was also confirmed by scanning tunneling microscopy (STM) (Fig. 2(b)). Steps of single BL were clearly observed. The step height is $\sim$ 3.95$\pm$0.05 \AA. The distance between two adjacent lines in the RHEED pattern is inversely proportional to the in-plane lattice constant. Fig. 2(c) presents the in-plane lattice constant of Bi ($a_{Bi}$) as a function of the film thickness. Below about 13nm, $a_{Bi}$ increases from 4.38 \AA (the same of the  Bi$_2$Te$_3$ substrate) to 4.54 \AA (bulk value of Bi). After that, $a_{Bi}$ does not change anymore. In this work, 20 nm and 30nm films were studied. X-ray diffraction (XRD) measurements were carried out on 30 nm Bi(111)/40 nm Bi$_2$Te$_3$/Si(111) and 30 nm Bi(111)/Si(111). Shown in the insert of Fig. 2(d), Bi's diffraction peaks in both films are almost coincide. The peak width of the Bi(003) peak in Bi(111)/Bi$_2$Te$_3$(111) is slightly smaller than on Si(111) (Fig. 2(d)), which implies that we may have better quality Bi(111) film on Bi$_2$Te$_3$ using our two-step growth method.

\begin{figure}[]
\center \includegraphics[width=8cm]{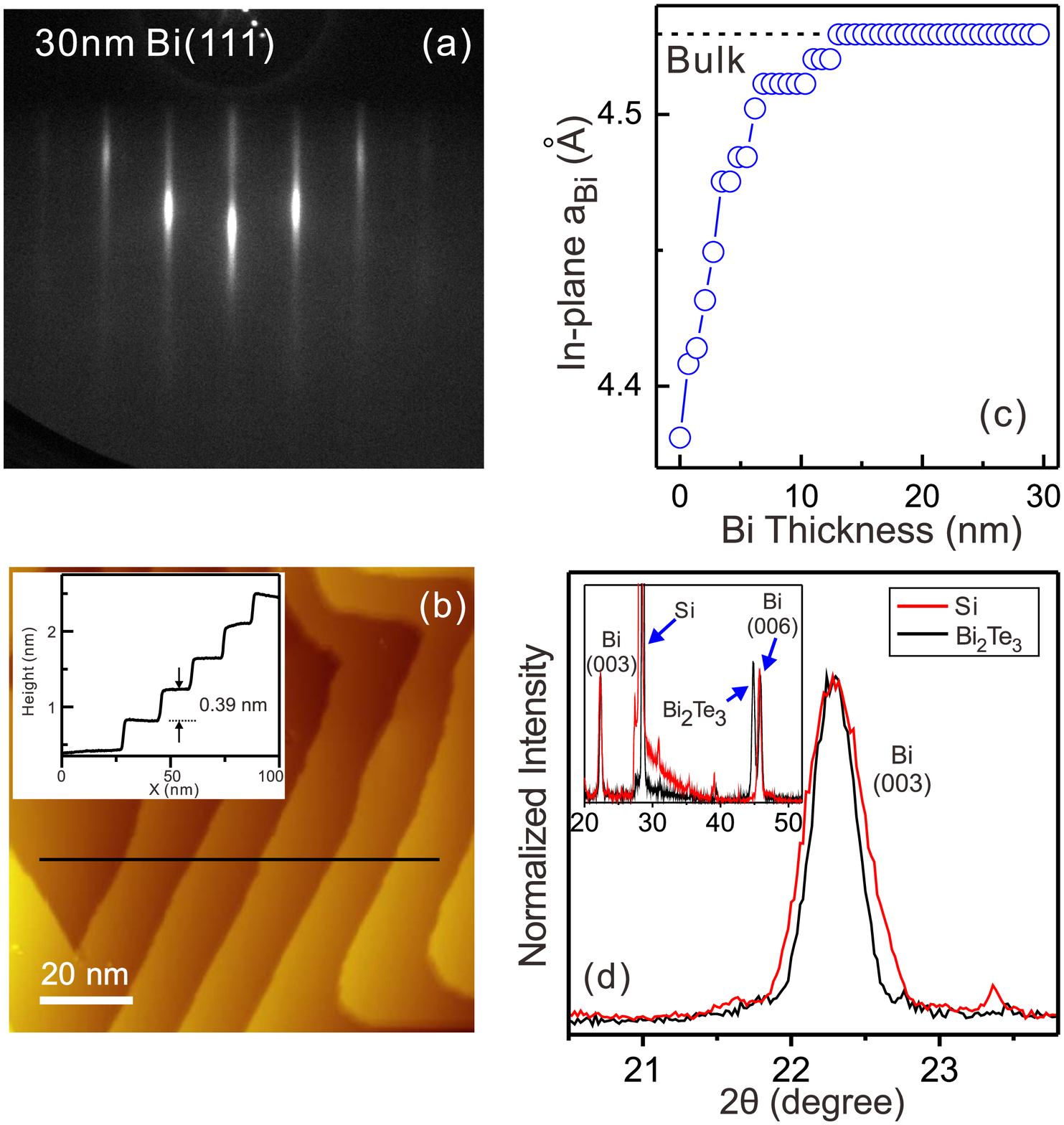} \caption{(a) RHEED pattern and (b) STM topograhpy of 30 nm Bi(111) films on Bi$_2$Te$_3$(111). (c) In-plane lattice constant of Bi films as a function of the thickness. (d) XRD spectra of 30 nm Bi(111)/40nm Bi$_2$Te$_3$/Si(111) (black line) and 30 nm Bi(111)/Si(111) (red line).}
\end{figure}

\begin{figure}[t]
\center \includegraphics[width=9cm]{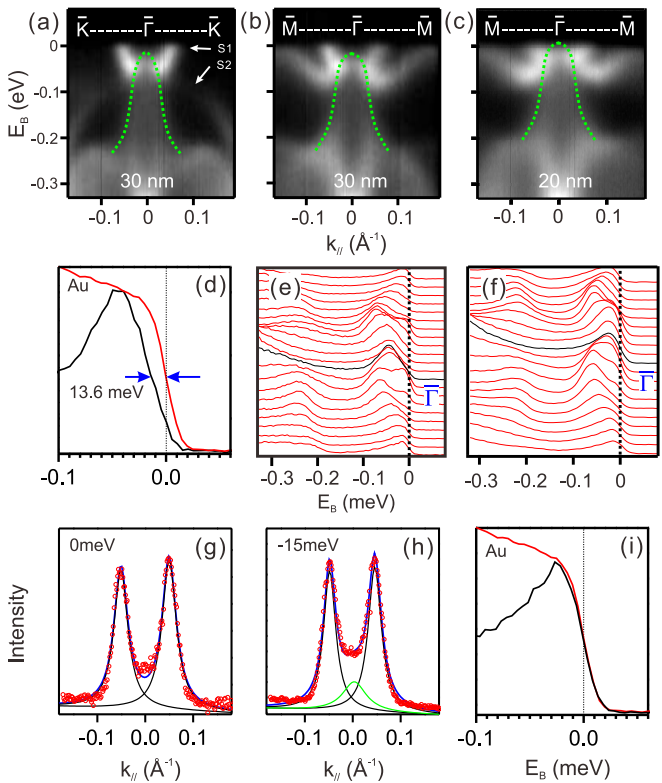} \caption{(a) and (b) ARPES spectra of 30 nm Bi(111)/Bi$_2$Te$_3$(111) along $\bar{\Gamma}$-$\bar{M}$ and along $\bar{\Gamma}$-$\bar{K}$ directions near the $\bar{\Gamma}$ point taken using 30 eV photons. The green dash lines mark the bulk valence band. (c) ARPES spectra of 20 nm Bi(111) along $\bar{\Gamma}$-$\bar{M}$ direction taken using 30 eV photons. (d) Black line is the the EDC of 30nm Bi at the $\bar{\Gamma}$ point. Red line is the EDC from the polycrystalline Au. (e) and (f) EDCs corresponding to (a) and (c). MDC curves from 30 nm Bi at (g) Fermi level and (h) E$_B$=-15 meV. Red dots are the experimental data and blue curves are the fitting results. At Fermi level, MDC can fitted by two Lorentzian peaks (black curves). At E$_B$=-15 meV, MDC can only fitted by at least three Lorentzian peaks (black and green curves).}

\end{figure}

\begin{figure*}[t]
\center \includegraphics[width=17cm]{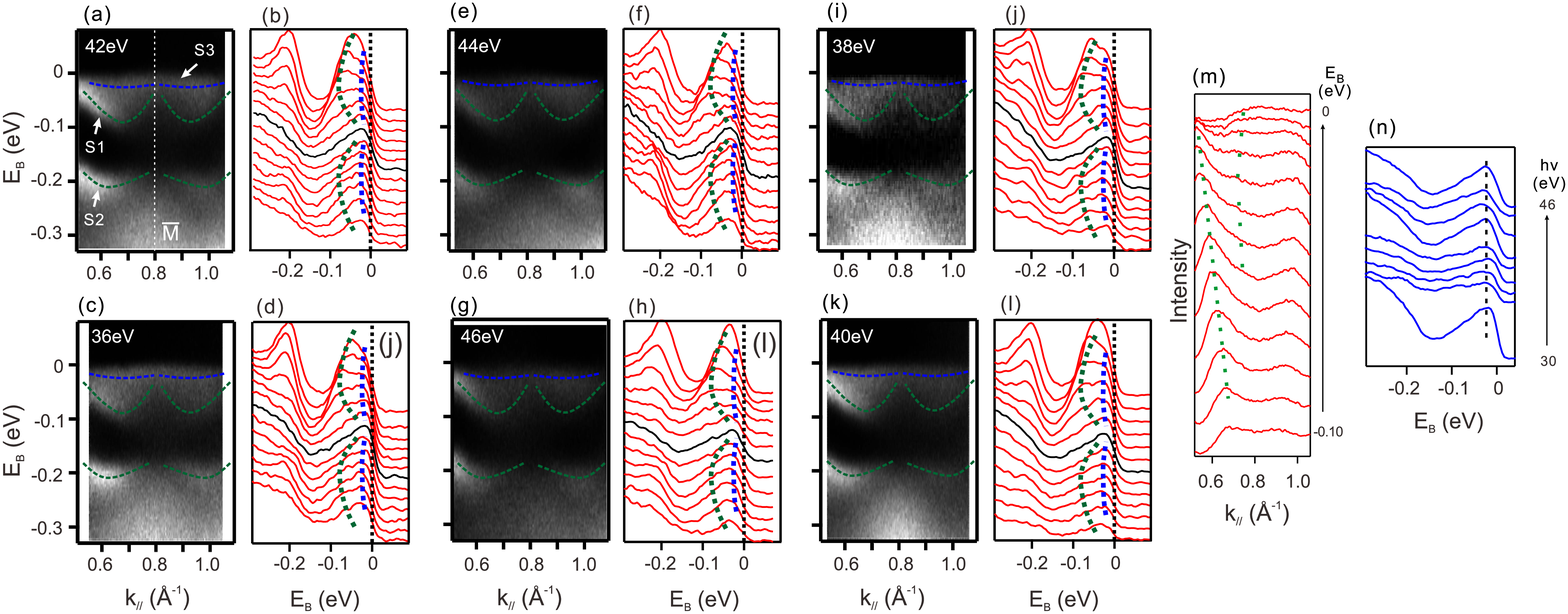} \caption{(a-l) High resolution ARPES spectra near the $\bar{M}$ point along the $\bar{\Gamma}-\bar{M}-\bar{\Gamma}$ direction using different incident photon energies. Green lines mark the surface bands of S1 and S2 that are the same bands observed near the $\bar{\Gamma}$ point. Blue dashed lines mark the new observed surface band. The black curves in all EDC plots represent EDCs right at the $\bar{M}$ point. (m)MDCs near the Fermi level from (a). Greed dots mark the MDC peaks of the S1 band. (n) EDCs at the $\bar{\Gamma}$ point as a function of the incident photon energy. The S3 band does not move.}
\end{figure*}

Figure 3(a)-3(c) present the low energy ARPES spectra of 30nm and 20nm Bi(111) films around the $\bar{\Gamma}$ point along the high symmetry directions. Figure 3(e) and 3(f) are the corresponding energy distribution curves (EDC) of Figs. 3(a) and 3(c). Consistent with the previously works, two surface states are identified (labeled as "S1" and "S2"). They merge together at the $\bar{\Gamma}$ point and split when away from the $\bar{\Gamma}$ point. Very close to the $\bar{\Gamma}$ point, there is a hole band (green dotted line is a guide for the eys) that is the bulk valence band\cite{Takayama2012}. The band maximum of the hole band is very close to the Fermi level. Figure 3(d) shows the EDC right at the $\bar{\Gamma}$ point (black curve in Fig. 3(e)) compared with the EDC from the polycrystalline Au. Clearly, there is an energy gap at this photon energy. Estimated from the leading edge of the Bi's EDC, the energy gap is about 13.6$\pm$2.5 meV. It should be noted that this experimental gap is not the real gap of the bulk valence band because we only measure at a single k$_z$ point. Because this valence band can only be clearly resolved under few incident energy, we failed to determine the exact position of the bulk valence band by tuning the incident photon energy (change the k$_z$). Figs. 3(g) and 3(h) show two momentum distribution curves (MDCs) taken at the Fermi level and at the binding energy (E$_B$) of 15 meV below Fermi level, respectively. In Fig. 3(g), at the Fermi level, MDC can be nicely fitted by two Lorentzian peaks originated from the S1 band. In contrast, at E$_B$=-15 meV, MDC can only be fitted by three Lorentzian peaks. The third peak (green curve in Fig. 3(h)) comes from the hole band. We also did ARPES measurements with the same incident photon energy (30 eV) on 20 nm Bi(111) thin films. Shown in Fig. 3 (c), the Fermi level of the 20nm film is slightly lower than that in the 30nm films. Fig. 3(i) shows the EDC at the $\bar{\Gamma}$ point of 20nm films in comparison with Au. Different from the 30nm films, no energy gap was observed, which is consistent with the previous study on 40 BLs Bi(111) films ($\sim$ 16nm) on Si(111)\cite{Takayama2012}. The higher Fermi level in our 30nm films provides us a good opportunity to carefully check the electronic structures at $\bar{M}$ point. Later on, we focused on the electronic structure of 30nm films near the $\bar{M}$ point.

Figure 4(a) - 4(n) show the low energy ARPES spectra and the corresponding EDCs of 30nm Bi(111) films near the $\bar{M}$ point along the $\bar{M}-\bar{\Gamma}-\bar{M}$ direction under different photon energies. Similar to the previous results on the bulk Bi(111)\cite{Ast2003,Hofmann}, the spectra near the $\bar{M}$ point are much weaker than those near the $\bar{\Gamma}$ point. In Fig. 4(a), two spin-split surface bands, S1 (part of the S1) and S2, are resolved (indicated by green dotted lines). S2 band is "W"-shape around the $\bar{M}$ point. Meanwhile, only part of the S1 band is clear where k $< $ 0.7 \AA$^{-1}$. In this momentum region, well separated peaks from the S1 band are observed in the EDCs (Fig. 4(b)). However it is difficult to trace the peaks of the S1 band in EDCs where k $> 0.7 \AA^{-1}$. In this region, we have to follow the intensities in the image plots (Fig. 4(a)) to get the possible dispersion of the band S1. The image plots present some properties of the momentum distribution curves (MDCs). Actually, part of the S1 band where k $> 0.7 \AA^{-1}$ can be traced in MDCs shown in Fig. 4(m). Green dotted lines in Fig. 4(m) mark the possible MDC peaks of S1 band. Due to the weak signals in MDCs and EDCs, the extracted dispersion relation of the S1 band has some uncertainty. Nevertheless, we observed a "W"-shape S1 band around the $\bar{M}$ point. The dispersion of the S1 band is also overlaid on the EDCs plots (Green dotted lines in EDC plots). On the other hand, very interestingly, we observed another band (labelled as band "S3", marked by the blue dashed line in Fig. 4(a)) besides the S1 and S2 bands. The S3 band is obvious in the EDCs (blue dashed line in Fig. 4(b)). This S3 band is very close to the Fermi level. Where does this S3 band come from? There are two possibilities. The first possibility is that the S3 band is the missing bulk conduction band of Bi. However, the S3 band is much flatter than the bulk conduction band in LDA calculations\cite{exp4}. The second possibility is that the S3 band is a surface state. To check this, we did the photon energy dependent ARPES measurements to change the k$_z$. The dispersion of the S3 band barely changes in Figs. 4(a) to 4(l). Figure 4(m) shows the EDCs right at the $\bar{M}$ point as a function of the incident photon energy. Though the intensity varies, the peak position of the S3 band does not change, which indicates that the S3 band is a surface state. The S3 band was only clearly observed on the 30nm films where we have highest Fermi level. According to the Kramers theorem, the S1 band must be degenerate with another surface band or merge into bulk band. In our films, the bulk conduction band is still missing in the ARPES spectra, however the third surface band S3 is observed. ARPES spectra in the Fig. 4 strongly suggest that the S1 band is degenerate with the S3 band at the $\bar{M}$ point.

Figure 5 presents the ARPES spectra of 30 nm Bi(111)/Bi$_2$Te$_3$ films from one time reversal invariant momenta point ($\bar{\Gamma}$) to another time reversal invariant momenta point ($\bar{M}$). Green dotted lines mark the observed surface bands. There are five Fermi crossing points indicated by the white arrows in the figure. Similar to the Bi$_{1-x}$Sb$_x$\cite{Hsieh2008}, odd number of Fermi crossing of the surface bands implies that 30 nm Bi(111)/Bi$_2$Te$_3$ is topologically nontrivial. It is worth to note that the physical reason why Bi(111)/Bi$_2$Te$_3$ is topological nontrivial can not be answered specifically from our experiments. It could be the intrinsic properties of Bi thin films of a certain thickness as the recent theory suggested\cite{Jinarxiv}. Or it could be unique in the Bi(111)/Bi$_2$Te$_3$ system. The in-plane lattice constant of Bi is compressed in the first 13 nm, which means Bi(111)/Bi$_2$Te$_3$ film is not exactly the same as the freestanding-like Bi(111) films. Recently calculations showed that bulk Bi can be a 3D topological semi-metal if the in-plane lattice is compressed\cite{Hirahara2012}.

\begin{figure}[]
\center \includegraphics[width=8cm]{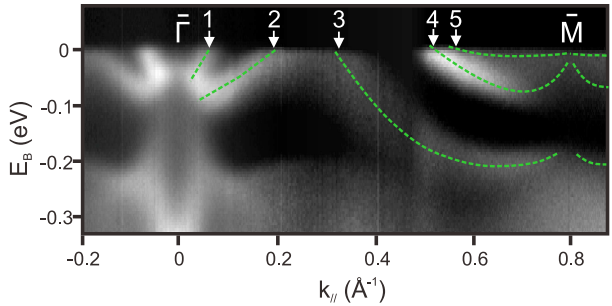} \caption{ARPES spectra from $\bar{\Gamma}$ point to $\bar{M}$ point. The white arrows indicate the Fermi crossing positions of the surface bands. Odd number of crossing points were observed.}
\end{figure}

In summary, we determined the electronic structure of 30 nm Bi(111) films grown on the Bi$_2$Te$_3$ substrate. By observation of the third surface band near the $\bar{M}$ point, we found the directly experimental signature that Bi(111)/Bi$_2$Te$_3$ film can be topologically nontrivial. The origin of the nontrivial properties in Bi(111)/Bi$_2$Te$_3$ films needs further investigation in theory in the future.

This work is supported by National Basic Research Program of China (Grants No. 2012CB927401, No. 2013CB921902), NSFC (Grants No. 11134008, No. 11174199, No. 11374206, No. 11274228, No. 11227404, No. 91421312, and No. 91221302), Shanghai Committee of Science and Technology (12JC140530 and No. 13QH1401500), C.L.G. acknowledges support from the Shu Guang project, which is supported by the Shanghai Municipal Education Commission and Shanghai Education Development Foundation. J.F.J. acknowledges support from the SRFDP and RGC ERG Joint Research Scheme of Hong Kong RGC and the Ministry of Education of China (No. 20120073140016, M-HKU709/12). D.Q. acknowledges support from the Top-notch Young Talents Program and the Program for Professor of Special Appointment (Eastern Scholar). The Advanced Light Source is supported by the Director, Office of Science, Office of Basic Energy Sciences, of the US Department of Energy under Contract No. DE-AC02-05CH11231.

\end{document}